# Ferromagnetism of sputtered $Fe_3GeTe_2$ ultrathin films in the absence of two-dimensional crystalline order

Qianwen Zhao[1,2], ChaoChao Xia[1,3], Hanying Zhang[1,2], Baiqing Jiang[1,2], Tunan Xie[1,2], Kaihua Lou[1,2], and Chong Bi[1,2,3*]

[1]State Key Lab of Fabrication Technologies for Integrated Circuits, Institute of Microelectronics, Chinese Academy of Sciences, Beijing 100029, China

[2]University of Chinese Academy of Sciences, Beijing 100049, China

[3]School of Microelectronics, University of Science and Technology of China, Hefei 230026, China

*bichong@ime.ac.cn

**Abstract:**

The discovery of ferromagnetism in two-dimensional (2D) monolayers has stimulated growing research interest in both spintronics and material science. However, these 2D ferromagnetic layers are mainly prepared through an incompatible approach for large-scale fabrication and integration, and moreover, a fundamental question whether the observed ferromagnetism actually correlates with the 2D crystalline order has not been explored. Here, we choose a typical 2D ferromagnetic material, $Fe_3GeTe_2$, to address these two issues by investigating its ferromagnetism in an amorphous state. We have fabricated nanometer-thick amorphous $Fe_3GeTe_2$ films approaching the monolayer thickness limit of crystallized $Fe_3GeTe_2$ (0.8 nm) through magnetron sputtering. Compared to crystallized $Fe_3GeTe_2$, we found that the basic ferromagnetic attributes, such as the Curie temperature that directly reflects magnetic exchange interactions and local anisotropic energy, do not change significantly in the amorphous states. This is attributed to that the short-range atomic order, as confirmed by valence state analysis, is almost the same for both phases. The persistence of ferromagnetism in the ultrathin amorphous counterpart has also been confirmed through magnetoresistance measurements, where two unconventional switching dips arising from electrical transport within domain walls are clearly observed in the amorphous $Fe_3GeTe_2$ single layer. These results indicate that the long-range ferromagnetic order of crystallized $Fe_3GeTe_2$ may not correlate to the 2D crystalline order and the corresponding ferromagnetic attributes can be utilized in an amorphous state which suits large-scale fabrication in a semiconductor technology-compatible manner for spintronics applications.

**INTRODUCTION**

Long-range ferromagnetic order which cannot persist down to two-dimensional (2D) regimes at a finite temperature has been proved in the well-known Mermin-Wagner theorem for several decades[1]. However, in recent years, many groups report that this restriction, arising from thermal fluctuations, can be counteracted by magnetocrystalline anisotropic fields (or external magnetic fields) that may open an energy gap in the dispersion of thermally excited magnons for stabilizing long-range magnetic order in the 2D regimes at a nonzero temperature[2–5]. Experimentally, ferromagnetism has been discovered in plenty of van der Waals materials down to a few layers or even the monolayer limit, such as $CrI_3$, $Cr_2Ge_2Te_6$ and $Fe_3GeTe_2$[6–8]. Like other conventional 2D materials, the typical approach to obtain these ferromagnetic low-dimensional materials is by using mechanical exfoliation of their bulk counterparts, which, nonetheless, is facile neither for large-scale production, nor for integration with other materials, especially in spintronic applications where the spin transport between two functional layers is extremely sensitive to the quality of interfaces[9] (the same situations for molecular beam epitaxy fabrications). More importantly, practical spintronic products, for example, magnetic sensors[10,11], hard disk drives[12–14], and magnetic random-access memory (MRAM)[15], do not require single crystalline or even polycrystalline magnetic materials, but instead, uniform amorphous magnetic materials that can be fabricated through the complementary metal oxide semiconductor (CMOS)-compatible techniques are preferred[10–16]. This is because, besides the compatibility of fabrication, the ferromagnetic properties of most conventional magnetic materials keep well in an amorphous state[17–21] (except crystal orientation-dependent anisotropy), and furthermore, no electrical noises or random domain-wall pinning sites formed at grain boundaries or along the crystal directions will benefit device reliabilities during magnetization manipulation[10–15]. Therefore, from the point of view of spintronic applications, it will be interesting to verify if the observed ferromagnetism in low-dimensional materials is actually related to the 2D crystalline order, which will provide a technology benchmark for utilizing their ferromagnetic properties in CMOS-compatible processes.

Theoretically, unlike electrical properties determined by the electronic band structure that strongly correlates with long-range crystalline order, magnetic properties mainly rely on short-range atomic order[16,18,19,21]. A simple yet successful model to describe magnetism is using a Heisenberg Hamiltonian

$$H = \sum_{i,j} J_{ij} \mathbf{S}_i \cdot \mathbf{S}_j + \sum_i D(\mathbf{n}_i \cdot \mathbf{S}_i)^2 + \mathbf{H} \cdot \sum_i \mathbf{S}_i, \qquad (1)$$

where $\mathbf{S}_i$ is the spin operator on site $i$, $\mathbf{n}_i$ is a unit vector of the local atomic anisotropy axis, $J_{ij}$ is the exchange coupling strength between sites $i$ and $j$, and H is the applied magnetic field. The first term describes the sum of isotropic Heisenberg exchange interactions between spin $i$ and neighbored spin $j$, and the second term describes the contribution from uniaxial magnetic anisotropy with an anisotropy constant $D$. Considering a material with fixed atomic ratios, in a solid crystalline state, both terms can be well calculated by using detailed lattice structures, while in an amorphous state, they can be evaluated through a disorder-modified $J_{ij}$ and $D$, where the modification reflects the change of relative spatial position between two neighbored spin sites[18,22]. $J_{ij}$ is usually dominated by the short-range exchange interaction arising from electrons' antisymmetric wave function governed by the Coulombic interaction[18,22,23] and can be estimated through the so-called Bethe-Slater curve[22,24] based on the interatomic spacing that does not change largely in a stable equilibrium state. Therefore, $J_{ij}$ can be treated as $J_{ij} = \langle J_{ij} \rangle + \Delta J_{ij}$ with an average exchange strength $\langle J_{ij} \rangle$ and an exchange fluctuation $\Delta J_{ij}$ in amorphous states[22]. Indirect exchange interactions like super- or double-exchange mediated by non-magnetic atoms may be treated in the same way. Meanwhile, $D$ can also be represented by a "local field" in amorphous materials with a correlation length of several angstroms, where the sign and strength of $D$ are determined by spin-orbit interactions[16,18,22,23,25–28]. In some cases, stress, shape, or interface may also contribute $D$[18,22]. Given the fact that the third term describing the Zeeman interaction does not depend on crystallinity, the basic magnetic properties such as the Curie temperature ($T_c$) and saturation magnetization do not change remarkably in the crystalline and amorphous states for plenty of ferromagnets[17–22,25], and the corresponding experimental results can also be well explained by using Eq. (1) in both states[16,22,28]. Therefore, in principle, the magnetic attributes of layered van der Waals materials could also not correlate their crystalline order strongly, even more weakly than 3D ferromagnetic materials, by considering that the nearest-neighboring spin sites are limited in the 2D layer. So far, intensive efforts have been focused on the exfoliation and characterization of 2D magnetic materials[6–8], but the correlation between magnetism and 2D crystalline order has not been discussed.

In this work, we choose a widely investigated 2D ferromagnet, $Fe_3GeTe_2$ (FGT)[4,5], to explore the possible ferromagnetism in an amorphous state. By comparing to the reported ferromagnetic properties of crystallized FGT (c-FGT), deep insights into the role of 2D crystalline order on the observed magnetic properties can be gained. C-FGT belongs to the hexagonal crystal system, where the Te-Fe$_3$Ge-Te slabs lying in the *ab* plane stack along the *c* axis, coupled via vdW interaction. The $T_c$ of bulk c-FGT is about 220 K, which reduces to 130 K (may vary with substrates) when the thickness is down to the monolayer limit (0.8 nm)[4,5]. The ferromagnetic order persisting in a monolayer was attributed to the sustaining perpendicular magnetic anisotropy (PMA) that suppresses the thermally excited magnons[2–5,7]. These clear ferromagnetic behaviors of c-FGT will be helpful for comparison with its amorphous counterpart in this work. We prepared the amorphous FGT (a-FGT) thin films through magnetron sputtering from a FGT target under high vacuum conditions. The sputtered a-FGT was controlled between 1 nm and 120 nm in thickness (shortened for FGT(t) with t being the thickness in nm) and capped with a 2.5 nm $TaO_x$ or 10 nm $Si_3N_4$ layer unless otherwise specified, which was then patterned into a Hall bar structure for Hall and magnetoresistance measurements.

**RESULTS AND DISCUSSION**

*Structural Characterization.* We first characterized structural and compositional properties of sputtered FGT by using high-resolution transmission electron microscopy (HRTEM), scanning transmission electron microscopy (STEM) equipped with energy dispersive x-ray spectroscopy (EDS), and X-ray photoelectron spectroscopy (XPS). No capping layer was deposited for these samples so that the degree of natural oxidation under ambient conditions can also be evaluated. Figure 1a shows the HRTEM images of 30 nm FGT, in which a naturally oxidized surface up to 4.1 nm can be observed clearly, highlighting the essentials of $TaO_x$ or $Si_3N_4$ capping layers for ultrathin FGT samples. The roughness of surfaces or interfaces is less than 0.4 nm even for the 30 nm FGT, evidencing that the sputtered FGT can be continuous thin films when the thickness is close to the c-FGT monolayer limit (0.8 nm). The surface roughness of several nanometer-thick FGT has also been confirmed by atomic force microscopy (AFM), where the average surface roughness is about 0.21 nm for 1 nm FGT (see Supporting Information Figure S1). In a similar amorphous ferromagnetic layer, we have also demonstrated that the thickness of continuous thin films can be safely controlled down to 0.6 nm through the magnetron sputtering[29]. The structural and compositional distributions have also been examined in a region of 30 nm × 1 μm, and no crystalline lattices or element aggregates were found. The typical structural and elemental mapping images shown in Figure 1a,b provide direct evidences that the sputtered FGT thin films are amorphous (see Supporting Information Figure S2 for X-ray diffraction results) and uniform in structure and composition.

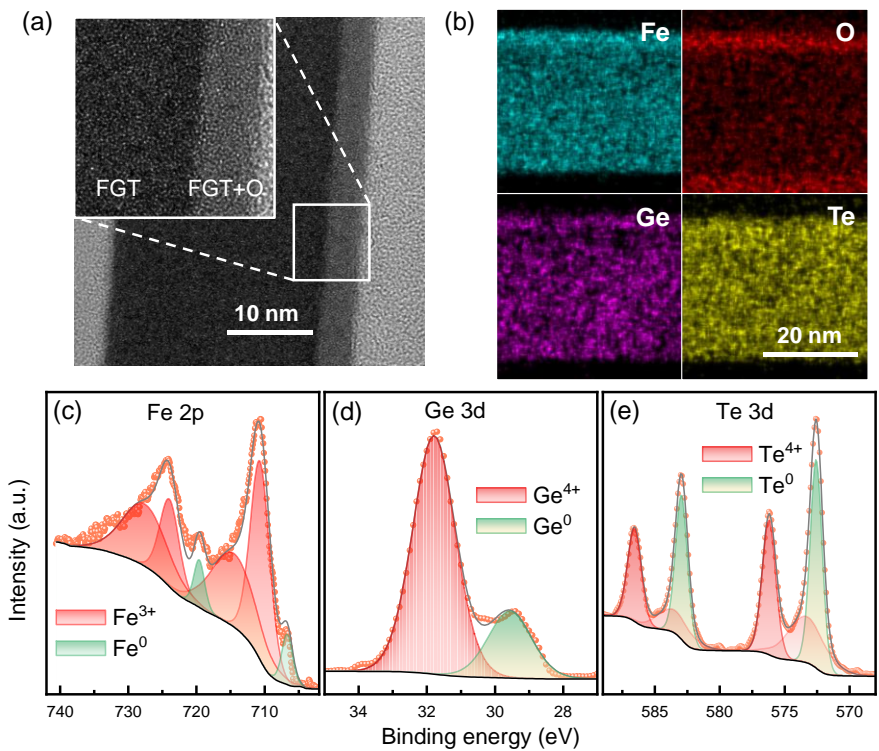

Figure 1. (a) HRTEM images of sputtered FGT thin films with the thickness of 30 nm. The enlarged part shows that the naturally oxidized layer is about 4.1 nm near the top surface. (b) Corresponding Fe, O, Ge, and Te maps of the 30 nm FGT. (c-e) XPS spectra of Fe 2p, Ge 3d and Te 3d. The filled peaks with different colors represent different valence states of each element. The isolated dots and connected lines are experimental data and fitting results, respectively.

The most important feature of sputtered FGT that can show similar magnetic properties as c-FGT is the valence states of each element, which directly reflect the chemical bond and short-range atomic order and determine the strength of direct exchange as well as spin-orbit coupling[18,22]. To analyze the valence bond states, we performed XPS measurements, as shown in Figure 1c-e. Before the XPS data was collected, the top oxidized surface had been etched. Generally, the XPS spectra of each element in c-FGT can be well reproduced in the sputtered FGT[30,31]. As shown in Figure 1c, the Fe 2p spectrum can be deconvoluted into six peaks, where the peaks at 706.6 eV/719.8 eV attributed to $Fe^0$ and 710.9 eV/724.6 eV attributed to $Fe^{3+}$ are exactly the same as those of crystallized bulk FGT. Two other peaks at 713.2 eV/ 726.2 eV can be attributed to the satellites of 710.9 eV/724.6 eV peaks. For the Ge 3d and Te 3d spectra shown in Figure 1d,e, the deconvoluted peaks are also the same as those of c-FGT[31]. These XPS results provide strong evidences that the valence bond states and local positions of each elemental atoms in c-FGT, at least in the scale of nearest-neighbors, are also sustained in the sputtered a-FGT. Correspondingly, those attributes mainly relying on the nearest-neighboring interaction, such as magnetism, may also maintain in the sputtered a-FGT. It should be noted that the detailed chemical states of each element determined through XPS spectra have not been well understood even in c-FGT. For instance, the peak around 724.6 eV was also attributed to $Fe^{2+}$ in some works[30], while the $Fe^{3+}$ signals were explained as originating from surface oxidation[31] but still appear in the samples after removing oxidized top surfaces[30]. Regardless of these debates on the origin of detailed peaks, the identical XPS results between c-FGT and sputtered a-FGT demonstrate that the same chemical states of each element are sustained in both phases. The estimated atomic percentages of Fe, Ge, and Te from XPS spectra are 42.1%, 24.6%, and 33.3%, respectively, in which the Fe concentration is less while Ge is higher than their corresponding concentrations in FGT sputtering targets with the same components as c-FGT (Fe decreases from 50.0% to 42.1%, and Ge increases from 16.7% to 24.6%). The low Fe concentration was also confirmed by using energy dispersive X-ray spectrometry (EDX; see Supporting Information Table S1). Compared to c-FGT, the lower Fe concentration of sputtered c-FGT indicates that there are enough Ge and Te to form Ge-Fe or Te-Fe bonds and probably no extra isolated Fe atoms contributing to the ferromagnetism as discussed below.

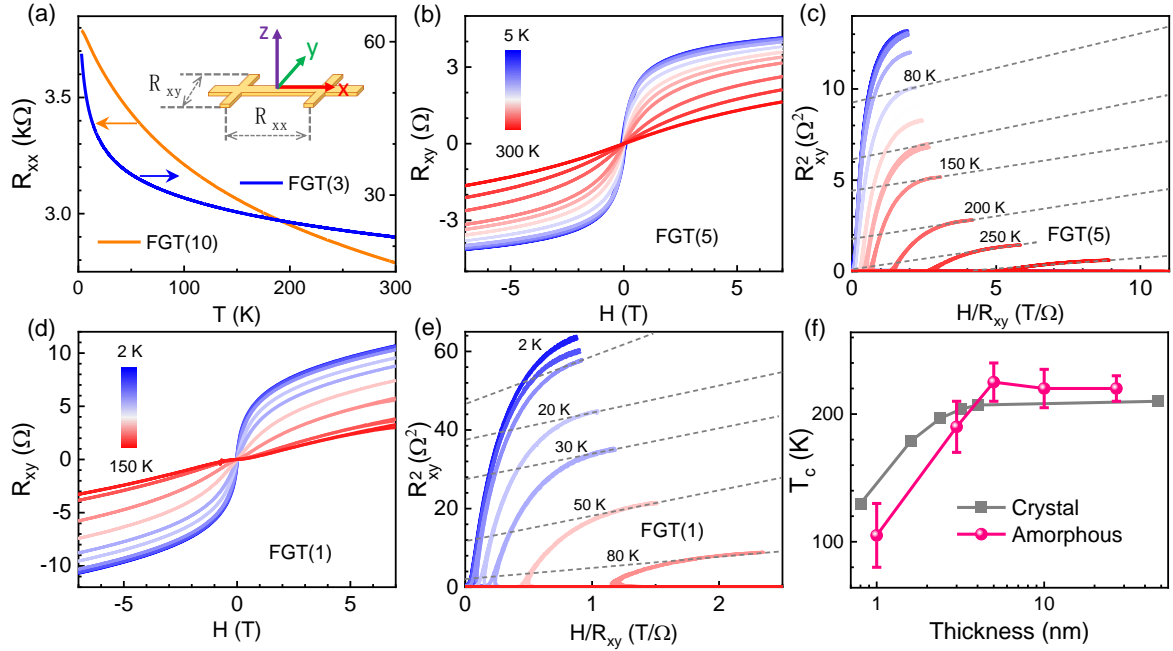

Figure 2. (a) $R_{xx}$ of sputtered a-FGT as a function of temperature with two typical thicknesses. Inset schematically shows the experimental configuration for electrical measurements. (b-e) $R_{xy}$ curves (b, d) and corresponding Arrott plots (c, e) of 5 nm (b, c) and 3 nm (d, e) a-FGT measured at different temperatures under a perpendicular magnetic field. (f) The thickness dependence of extracted $T_c$ for sputtered a-FGT. The $T_c$ data of c-FGT is from other references[5] for comparison.

*Determination of $T_c$ through Hall Measurements*. The longitudinal resistance ($R_{xx}$) and anomalous Hall resistance ($R_{xy}$) of sputtered FGT were examined by using a four-point measurement configuration, as schematically shown in the inset of Figure 2a. The temperature dependences of $R_{xx}$ shown in Figure 2a illustrate that the sputtered FGT shows semiconducting behaviors for all thicknesses up to 60 nm, in sharp contrast to c-FGT that is metallic in bulk and becomes semiconducting down to trilayer (2.4 nm). This can be understood that the electrical transport properties of both a-FGT and c-FGT thinner than trilayer are dominated by disorder, mainly interlayer (or interfacial) structural disorder in the trilayer or thinner c-FGT, although in which the intralayer 2D crystalline order still keeps. Figure 2b and Figure 2d present $R_{xy}$ curves under a perpendicular magnetic field ($H_z$) for 5 nm and 1 nm FGT, respectively. At 300 K, only the linear response contributed from the normal Hall effects can be observed, indicating that there is no ferromagnetism in the sputtered FGT at room temperature. With decreasing temperature, the typical ferromagnetic hysteresis appears for both samples and finally dominates $R_{xy}$ signals at low temperatures below 100 K. The transition temperature to ferromagnetism strongly depends on the thickness of sputtered FGT like c-FGT and other ferromagnetic materials.

To determine $T_c$ accurately, the Arrott plots[32] of $R_{xy}$ curves for 5 nm and 1 nm a-FGT, $R_{xy}^2$ as a function of $H/R_{xy}$, are replotted in Figure 2c and Figure 2e, respectively, where the positive intercept of linear part in the high-field range can be thought as ferromagnetic states. $T_c$ is determined when the intercept of linear part approaches zero. For the thickness larger than 5 nm, the determined $T_c$ is about 225 K, which is close to that of c-FGT by considering around 25 K misestimation due to the interference of normal Hall contribution. As shown in Figure 2e, for 1 nm a-FGT, the temperature at which the intercept approaches zero is apparently lower than that of 5 nm FGT because of a reducing $T_c$. The thickness dependences of $T_c$ for c-FGT[4,5] and sputtered a-FGT are shown in Figure 2f, in which similar $T_c$ values for both phases emerge at each thickness. These results indicate that the basic exchange interactions responsible for the observed ferromagnetism, which are directly reflected through $T_c$, are not changed extraordinarily in both phases. Therefore, the fundamental magnetic attributes of c-FGT dominated by the exchange interactions may not be related to the long-range crystalline order, but instead, are mainly determined by the short-range atomic order like electron transport behaviors down to trilayer (as discussed above in Figure 2a). Remarkably, for the 1 nm a-FGT approaching the thickness limit of c-FGT monolayer, ferromagnetic characteristics still keep well below 100 K as shown in Figure 2d, indicating the existence of ferromagnetism.

*Demonstration of Ferromagnetism through Magnetoresistance Measurements*. Previous studies have attributed the ferromagnetism of c-FGT in a few layers to the appearance of PMA suppressing magnon excitation[4,5]. As shown in Figure 2b,d, the sputtered a-FGT does not show PMA, but does show hysteresis behaviors down to 1 nm at low temperatures. One possible reason for stabilizing ferromagnetism in the sputtered a-FGT can be the in-plane magnetic anisotropy (IMA)[26]. To reveal IMA and further confirm ferromagnetism, we performed anisotropic magnetoresistance (AMR) measurements[33] by detecting angle dependence of $R_{xx}$ within different planes, as schematically shown in the inset of Figure 3a. Moreover, AMR measurements can also provide clear evidences to distinguish possible superparamagnetism induced by magnetic nanoparticles or aggregates, which is always isotropic as demonstrated in granular films[34,35]. The applied current is along the *x* direction and the strength of applied magnetic field (H) was fixed at 6 T, which is large enough to saturate magnetization in all directions as demonstrated in Figure 2b,c. Figure 3 and Figure 4 present the typical AMR results of 3 nm a-FGT at several representative temperatures. The absence of magnetic field and angle dependences of $R_{xx}$ at 300 K are consistent with $R_{xy}$ results (Figure 2b), further confirming no ferromagnetism at room temperature.

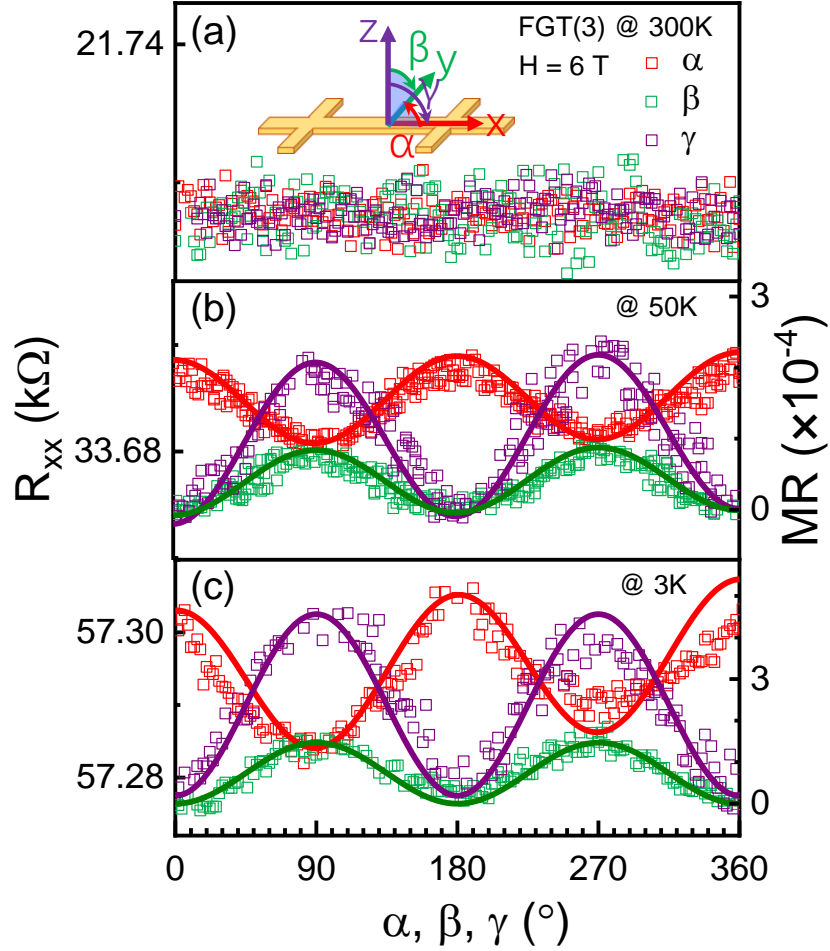

Figure 3. (a-c) The angle dependences of $R_{xx}$ for 3 nm a-FGT at 300 K (a), 50 K (b), and 3 K (c). The solid lines are $\sin(2\varphi)$ or $\cos(2\varphi)$ fitting results, where φ represents α, β, or γ in each scan plane as defined in the inset of (a). The applied external field is 6 T.

When the temperature drops to 50 K, as shown in Figure 3b, $R_{xx}$ shows clear angle dependences. According to the AMR theory, $R_{xx}$ depends on the angle (φ) between electrical current and magnetization, that is, $R_{xx} \propto cos^2\varphi$ and shows maximum when magnetization is along to the current direction. For the α and γ scans shown in Figure 2b, $R_{xx}$ can be explained through the AMR mechanism. However, the β scan results contradict AMR theory since $R_{xx}$ should not change when magnetization is perpendicular to current and the angle dependence in the β scan is not expected[33]. The β dependent $R_{xx}$ is usually observed in a ferromagnetic multilayer (usually a bilayer) involving a spin Hall layer or a Rashba interface due to spin Hall magnetoresistance (SMR)[36–39], or in a thin ferromagnet due to geometrical size effects (GSE)[40,41]. SMR appears when the spin polarization generated through the spin Hall or Rashba effects (along the *y* direction) modulates the spin absorption at the spin Hall or Rashba interfaces, which induces a resistance change when the spin polarization is parallel or perpendicular to magnetization. In this case, the minimum of $R_{xx}$ corresponds to H along the *y* direction. This does not agree with experimental data with a maximum value around β = 90º even though we consider that a spin current may be generated in a single ferromagnet with broken inversion symmetry[42–44]. Therefore, we attribute the β scan results to the GSE-related magnetoresistance, in which the β dependence of $R_{xx}$ is also consistent with that observed in most ferromagnetic materials[40]. As shown in Figure 3c, the AMR effects become more pronounced at 3 K and the corresponding magnetoresistance ratio (MR) of γ scan increases about two times, as expected originating from ferromagnetic behaviors which are usually enhanced with decreasing temperature.

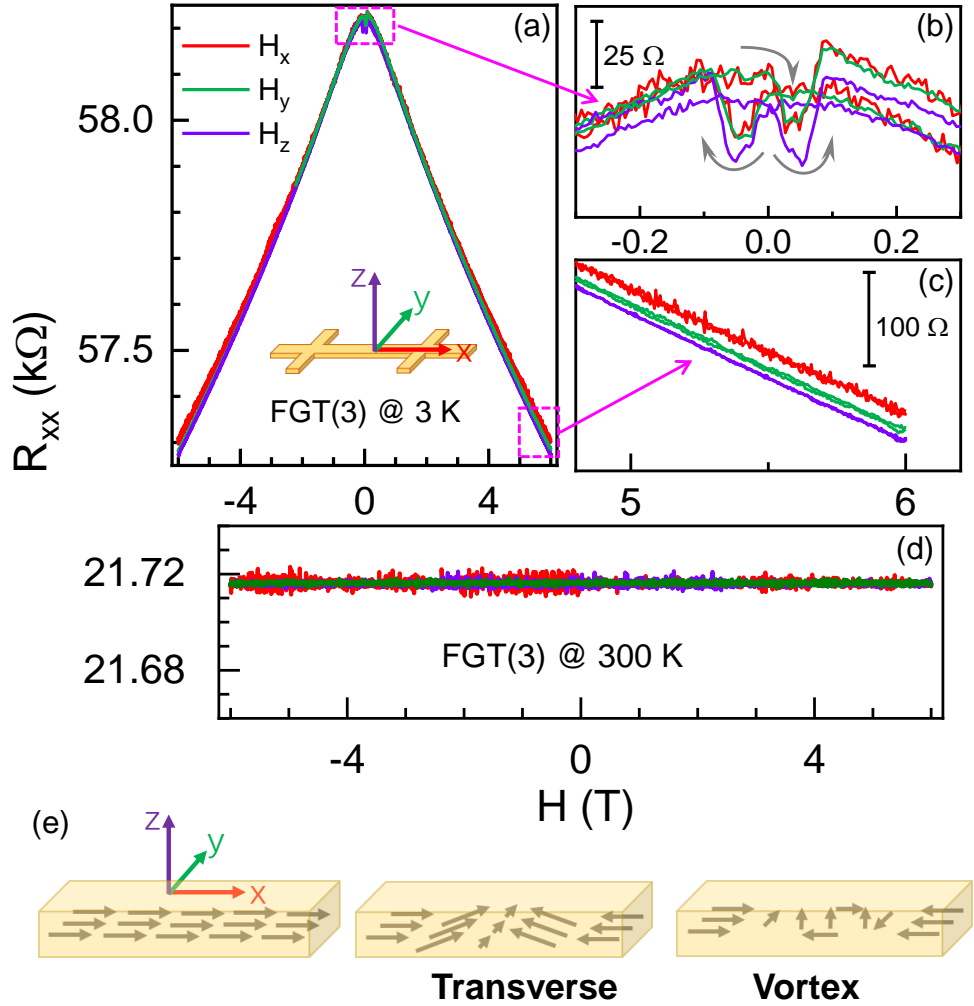

Figure 4. (a-c) $R_{xx}$ of 3 nm a-FGT as a function of applied external fields measured at 3 K. Inset schematically shows field directions. The enlarged parts in the low and high field ranges are shown in (b) and (c), respectively. The arrows in (b) indicate field sweep directions. (d) The corresponding $R_{xx}$ versus applied magnetic field at 300 K. (e) Schematic of transverse and vortex DWs.

To further examine the detailed magnetization switching driven by an applied magnetic field, $R_{xx}$ as a function of H along different directions was also recorded. Figure 4 shows the field-dependent $R_{xx}$ curves at two typical temperatures showing ferromagnetic or nonmagnetic characteristics respectively. At 3 K, a negative MR that $R_{xx}$ decreases with increasing H can be observed for all directional fields, as shown in Figure 4a. The negative MR has been demonstrated in batch of 2D materials with strong disorder[45–47], 2D electron systems[48,49], and ultrathin ferromagnetic layers[50–52]. The decreased resistance at high magnetic fields can be attributed to the reduced scattering probabilities of electrons from thermally excited magnons or short-range disorder such as ionized impurities[40]. In the sputtered a-FGT showing ferromagnetism, both the magnon and disorder scatterings occur and thus may contribute the negative MR simultaneously. Figure 4b and 4c show the enlarged $R_{xx}$ curves in low and high field ranges, respectively. At high fields, $R_{xx}$ curves do not overlap under different directional magnetic fields, reflecting the AMR effects shown in Figure 3c.

More interesting results are shown in Figure 4b, in which there are two dips, instead of two sharp peaks like conventional ferromagnets[33], appearing in the $R_{xx}$ curves around zero field. The two dips are the typical giant magnetoresistance (GMR) phenomena in spin valves or domain wall (DW) magnetoresistance in narrow ferromagnetic wires[53]. The former requires at least two ferromagnetic layers[9], which probably does not happen in a single ferromagnetic layer. The latter happens because the magnetization in DWs rotates to other directions and results in a resistance change. For example, as schematically shown in Figure 4e, for an in-plane magnetized ferromagnet along the *x* direction, the magnetization in a transverse DW will rotate to the y direction (partially to the *z* direction for vortex DWs)[54]. Since $R_{xx}$ (H//y) < $R_{xx}$ (H//x) according to the AMR theory, the resistance of DW regions and thus the total $R_{xx}$ reduces when multidomain states are formed during magnetization switching[53]. To verify the possible DW-related resistance change as the mechanism of two resistance dips in Figure 4b, we first inspect the field range where the two dips appear. As shown in Figure 4b, the two dips happen to appear in the field range where the multidomain state appears (180 Oe $\leq |H| \leq$ 750 Oe, also confirmed by direct magnetometry measurements shown below in Figure 5a), and thus, the reduced $R_{xx}$ can only be explained by considering the electrical transport within DWs. Second, the dips at positive and negative fields almost overlap for the field along *x* and *y* directions and the resistance drops about 28 Ω. The relative resistance change is about $4.81 \times 10^{-4}$, which is larger than the MR value of α and β scans but smaller than γ scans as shown in Figure 3c. This indicates that the magnetization of DWs in the multidomain state mostly rotates from the *x* to *z* direction (vortex DWs) and the resistance decrease is also consistent with Figure 3c where $R_{xx}$ (H//z) < $R_{xx}$ (H//x). For the $R_{xx}$ versus $H_z$ curve, the resistance drop is about 31 Ω and the relative resistance change (about $5.32 \times 10^{-4}$) is still smaller than the MR value of γ scan, which can also be understood as the DW-induced resistance decrease. Third, since the two dips in the three directional $R_{xx}$ curves can only be explained as the magnetization rotating from the *x* to *z* direction, it indicates a strong IMA along the *x* direction with the strength larger than 180 Oe (domain nucleation field). This is because, if the in-plane anisotropic field is smaller than 180 Oe, the magnetization should be aligned to the external field direction before domain nucleation and the dips in the $H_y$ and $H_z$ curves cannot arise from the *x* to *z* magnetization rotation. The IMA along *x* direction is also consistent with the expected shape anisotropy along the length of Hall bar.

It should be noted that, the two dips are in general the same as two switching peaks in the AMR curves of conventional ferromagnets except that the multidomain states survive in a large field range and the switching dips are negative for all directional fields. The negative dips can be understood that $R_{xx}$ in the multidomain states is dominated by electrical transport within the DW regions that result in a decreasing $R_{xx}$ as explained above. In conventional ferromagnets, $R_{xx}$ of the multidomain states is usually dominated by electrical transport within the domain (rather than DW) regions, where $R_{xx}$, and thus the sign of two switching peaks, is determined by the relative orientation between applied current and the total magnetization of all domains as predicted by the AMR theory[33]. The DW transport-dominated $R_{xx}$ in a single a-FGT layer not only confirms the persistence of ferromagnetism but also indicates the possible unconventional DW behaviors that may be interesting in theory and application. The direct observation of DWs requires low-temperature magneto-optical Kerr-effect (MOKE) microscope with high-spatial resolution, which is beyond the scope of current work. At 300 K, no any field-dependent $R_{xx}$ signals are detected due to the lack of ferromagnetism, as shown in Figure 4d.

*Magnetization Characterization through VSM Measurements*. The magnetization of sputtered a-FGT was also directly characterized by using vibrating sample magnetometer (VSM). To gain clear magnetic signals, a 10 nm a-FGT film was adopted for the VSM measurements. Figure 5a shows magnetization as a function of applied in-plane H at 50 K and 3 K, in which typical hysteresis loops due to ferromagnetism can be observed. Both the saturation magnetization and coercivity increase with decreasing temperature from 50 K to 3 K, in consistent with ferromagnetic behaviors in most conventional ferromagnets. At 3 K, the magnetization gradually switches to a reversed direction around zero field due to domain formation and the coercivity is about 310 Oe. The magnetization switching process and switching fields agree well with that revealed by AMR measurements shown in Figure 4b. Figure 5b shows the temperature dependence of magnetization under a 500 Oe in-plane H. The increase of magnetization with decreasing temperature further confirms the ferromagnetic behaviors. The temperature at which magnetization drops to zero is about 240 K, very close to $T_c = 220$ K determined through Hall measurements (Figure 2f). These VSM results provide direct magnetometry evidences for the appearance of ferromagnetism in the sputtered a-FGT.

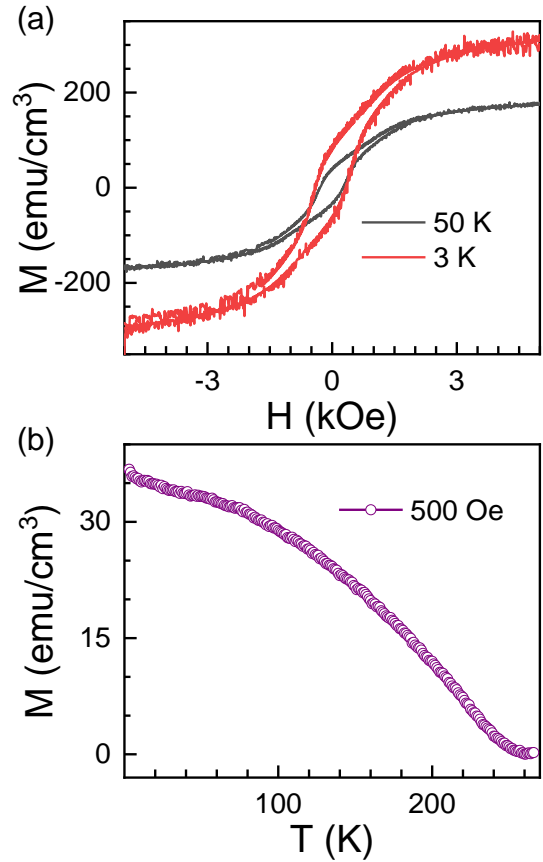

Figure 5. (a) In-plane magnetized hysteresis loops measured at 50 K and 3 K. A diamagnetic linear background due to substrates has been subtracted. (b) The temperature dependence of magnetization under a 500 Oe in-plane field.

*Discussion*. Ferromagnetism in 2D materials has attracted growing attention in both material and spintronic research areas. In fact, ferromagnetism in atomically thin conventional ferromagnets, such as Fe, Co, and Ni, has been investigated for several decades[55]. These materials are epitaxially grown on single crystal substrates to form special crystal orientations, and similar to ferromagnetic 2D materials, magnetocrystalline anisotropy is thought to be a key to removing the restriction of Mermin-Wagner theorem. However, as a fundamental question, the correlation between observed ferromagnetism and 2D crystalline order has never been explored. As mentioned above, clarifying this question is not only helpful for understanding the mechanism of ferromagnetism in 2D materials but also important in application to evaluate if they can be deposited through CMOS-compatible technologies. Through structural, electrical, and magnetization characterizations, our results incontrovertibly demonstrate that the ferromagnetism can also persist in a-FGT down to 1 nm in the absence of crystalline order.

First, as shown in Figure 1a,b, there are no visible crystalline regions, nanoparticles, or Fe aggregates in the sputtered FGT, which was examined by HRTEM with a spatial resolution less than 1 nm. Moreover, the Fe concentration estimated from XPS spectra (at. 42.1%) is also much less than that of c-FGT (at. 50%), indicating that there are probably no extra isolated Fe atoms. Even though there are some Fe aggregates with the size less than 1 nm that cannot be distinguished by HRTEM, they should show superparamagnetic behaviors like magnetic ion-doped granular films. The superparamagnetic features are isotropic[34,35] and not consistent with AMR results shown in Figure 3b,c. In addition, the clear magnetoresistance switching signals around zero field driven by applied H (Figure 4a,b) cannot arise from superparamagnetism.

Second, XPS spectra show that the chemical valence states of each element are exactly the same as c-FGT and all valence bond states in c-FGT can also be found in the sputtered a-FGT. These valence state results demonstrate that the chemical bond and thus the relative position between two atomic sites remains at least in the length of next-nearest neighbors in the sputtered a-FGT. Remarkably, the $T_c$ of a- and c-FGT that is mainly determined by the strength of exchange coupling between two neighbored sites[18,22,23] is also correspondingly the same with the thickness larger than 5 nm, indicating that the long-range crystallize order, both in-plane and out-of-plane, may not be the main factors determining the exchange interactions and local magnetic anisotropic fields in the layered c-FGT. As shown in Figure 2f, for the FGT layer less than 5 nm, $T_c$ shows almost the same reduction for both phases, also confirming similar exchange interactions and local magnetic anisotropic energies in c- and a-FGT, although the former usually shows PMA while the latter shows IMA. It should be noted that PMA of sputtered FGT can also be expected by introducing an interfacial PMA like amorphous CoFeB with selected buffer and capping layers[29].

Third, the ferromagnetism of amorphous FGT persists down to 1 nm, approaching the thickness of c-FGT monolayer. According to the Hall measurements, $T_c$ of the sputtered 1 nm FGT is about 100 K, which is very close to the reported $T_c$ values of c-FGT monolayer[5], demonstrating that the similar exchange interactions and local magnetic anisotropic energies in a-FGT as c-FGT maintain even below 1 nm. As demonstrated by AMR (Figure 4b) and VSM (Figure 5a) measurements independently, the sputtered ultrathin a-FGT films show IMA and gradually magnetization switching around zero field due to domain formation. The IMA may be the source to counteract restriction of the Mermin-Wagner theorem and stabilize ferromagnetism in ultrathin a-FGT films[26].

**CONCLUSIONS**

The ferromagnetism of a-FGT thin films down to 1 nm has been demonstrated by using electrical and magnetometry measurements. Similar to c-FGT monolayers with PMA, our results show that ferromagnetism can also persist in the in-plane magnetized a-FGT with the thickness close to the c-FGT monolayer limit, where IMA may take a critical role in the stabilization of long-range ferromagnetic order by creating a magnon energy gap. The $T_c$ of a-FGT is the same as that of c-FGT when the thickness is larger than 5 nm and also shows close valves for a thinner FGT below 3 nm, indicating that the two fundamental factors contributing ferromagnetism, exchange interaction and local magnetic anisotropy, are similar for both amorphous and crystallized phases and may not relate to 2D crystalline order. The clear ferromagnetic switching of a-FGT with domain formation is also revealed by using magnetoresistance and VSM measurements independently. In addition, the DW-dominated magnetoresistance that usually appears in very narrow ferromagnetic wires is also observed in the single a-FGT layer and can only be explained by considering vortex-type DWs, indicating the possible unconventional magnetic domain behaviors in the a-FGT.

From the viewpoint of spintronic applications, similar magnetic attributes in crystallized and amorphous FGT indicate that a-FGT that can be fabricated facilely in large scales may be used for replacing c-FGT in most devices. For instance, it will be interesting to verify large voltage control and GMR effects by employing a-FGT[5,56,57]. Moreover, as mentioned above, fabrication of perpendicularly magnetized a-FGT by using selected adjacent layers and exploration of possible special domain or skyrmion structures[58–60] and corresponding spin-orbit torque switching[61,62] are also interesting in applications. Like c-FGT, $T_c$ of a-FGT may also be increased up to room temperature by modulating Fe concentrations[63].

## EXPERIMENTAL SECTION

The a-FGT thin films were sputtered from a $Fe_3GeTe_2$ target with the purity higher than 99.9% through DC magnetron sputtering. The base vacuum before sputtering was pumped down to $8 \times 10^{-9}$ Torr and the Ar pressure during sputtering was set to 2 mTorr. The thickness of sputtered FGT was controlled between 1 nm and 120 nm by using a deposition rate about 0.17 Å/s with the DC power of 15 W. A 2.5 nm Ta or 10 nm $Si_3N_4$ layer as the capping layer was then subsequently deposited, in which the 2.5 nm Ta would be naturally oxidized when the samples were transferred out from the vacuum chamber. All samples were deposited on silicon wafers with a 300 nm thermally oxidized $SiO_2$ layer. By using the standard photolithography and ion milling processes, the deposited thin films were then patterned into a Hall bar structure (as schematically shown in the inset of Figure 2a) with the width of 10 μm and length of 50 μm for the Hall ($R_{xy}$) and resistance ($R_{xx}$) measurements. A 30 nm FGT without capping layers were also deposited for structural and compositional characterization by using commercial XPS (ESCALAB 250Xi) and HRTEM (FEI Titan Themis 200 TEM) measurements. $R_{xx}$ and $R_{xy}$ were measured by using Physical Properties Measurement System (PPMS, Quantum Design) in the temperature range of 3 - 300 K.

## ASSOCIATED CONTENT

The Supporting Information is available free of charge at

Supplementary Note, Figure S1-3, and Table S1 present structure and composition of sputtered FGT analyzed by AFM, X-ray diffraction (XRD), and scanning electron microscopy-energy dispersive X-ray spectrometry (SEM-EDX).

## ACKNOWLEDGMENTS

This work is supported by the National Key R&D Program of China (Grant No. 2019YFB2005800 and 2018YFA0701500), the National Natural Science Foundation of China (Grant No. 61974160, 61821091, and 61888102), and the Strategic Priority Research Program of the Chinese Academy of Sciences (Grant No. XDB44000000).

**Supporting Information**

# Ferromagnetism of Nanometer-Thick Sputtered Fe$_3$GeTe$_2$ Films in the Absence of Two-Dimensional Crystalline Order: Implications for Spintronics Applications

Qianwen Zhao[1,2], ChaoChao Xia[1,3], Hanying Zhang[1,2], Baiqing Jiang[1,2], Tunan Xie[1,2], Kaihua Lou[1,2], and Chong Bi[1,2,3]*

[1]State Key Lab of Fabrication Technologies for Integrated Circuits, Institute of Microelectronics, Chinese Academy of Sciences, Beijing 100029, China

[2]University of Chinese Academy of Sciences, Beijing 100049, China

[3]School of Microelectronics, University of Science and Technology of China, Hefei 230026, China

*E-mail: bichong@ime.ac.cn

**Supplementary Note 1: Surface roughness of sputtered FGT**

The surface roughness of sputtered FGT has also been confirmed by atomic force microscopy (AFM). Figure S1 shows AFM images of sputtered 1 nm FGT without capping layer, from which the calculated average roughness is about 0.21 nm. Please note that the 1 nm FGT has been oxidized according to TEM analyses as mentioned in main text.

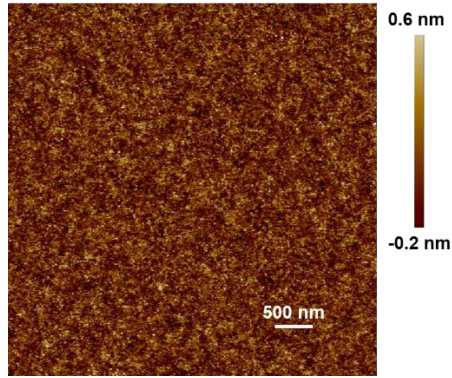

Figure S1. Surface morphology of 1 nm oxidized a-FGT thin films scanned by AFM.

**Supplementary Note 2: XRD results of sputtered FGT**

X-ray diffraction (XRD) measurements were performed by using a sputtered 120 nm FGT thin film without capping layers. No extra XRD peaks due to FGT can be detected, indicating no crystalline order formed in the sputtered FGT.

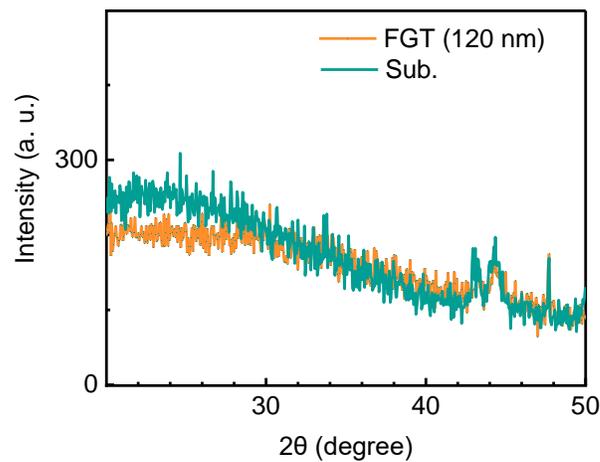

Figure S2. XRD results of sputtered 120 nm FGT (no capping layers). XRD signals from Si/SiO$_2$ substrates are also presented for comparison.

**Supplementary Note 3: Fe concentration of sputtered FGT analyzed by EDX**

Excessive Fe may induce ferromagnetism in the sputtered FGT. We have also measured the Fe concentration by using scanning electron microscopy-energy dispersive X-ray spectrometry (SEM-EDX). Figure S3 shows a typical cross-sectional SEM image of 60 nm sputtered FGT (no capping layers). EDX data was collected at two different positions, P$_1$ and P$_2$. The corresponding Fe, Ge, and Te concentrations are calculated and listed in Table S1, in which the measured Fe concentration approaches that acquired by XPS and is also much less than that of crystallized FGT.

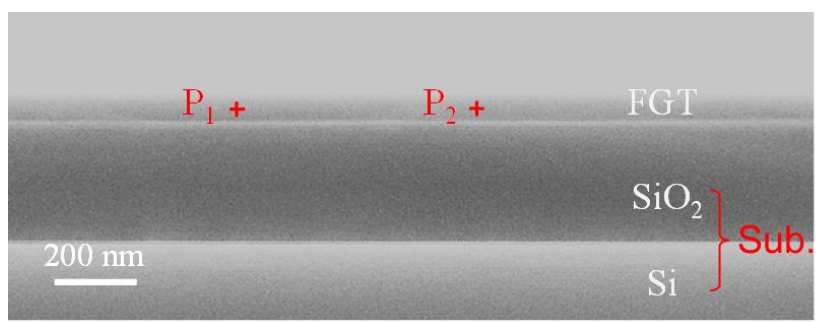

Figure S3. Cross-sectional SEM images of 60 nm sputtered FGT. P$_1$ and P$_2$ indicate two positions for acquiring EDX data.

Table S1. Composition of sputtered FGT acquired through SEM-EDX.

| Position | Fe (%, at.) | Ge (%, at.) | Te (%, at.) |
|---|---|---|---|
| P$_1$ | 38.09 | 28.62 | 33.29 |
| P$_2$ | 38.24 | 28.77 | 32.99 |